\documentclass[11pt,a4paper]{article}

\usepackage{jheppub}
\usepackage{xcolor}
\usepackage{lipsum}
\usepackage{ntheorem}
\usepackage{mdframed}

  {\begin{mdframed}[backgroundcolor=lightgray]\begin{mdtheorem}}%
  {\end{mdtheorem}\end{mdframed}}

\usepackage{graphicx}
\usepackage{xspace}   

\usepackage[utf8]{inputenc}

\newcommand{\ie}{{\it i.e.}\xspace}

\def\red{\color[rgb]{0,0,0}}

\def\comment{}

\newcommand{\be}{\begin{eqnarray}}
\newcommand{\ee}{\end{eqnarray}}
\newcommand{\eto}{{\rm e}}
\newcommand{\rhat}{\hat{r}}

\def\funp{{I\!\!P}}

\title{Saturation model of DIS: an update}
\author[a,b]{Krzysztof Golec-Biernat}
\author[a]{Sebastian Sapeta}

\affiliation[a]{Institute of Nuclear Physics, Polish Academy of Sciences, 31-342 Cracow, Poland}
\affiliation[b]{Faculty of Mathematics and Natural Sciences, University of Rzesz\'ow,  35-959 Rzesz\'ow, Poland}

\emailAdd{golec@ifj.edu.pl}
\emailAdd{sebastian.sapeta@ifj.edu.pl}

\abstract{We present the results of new fits to the recently extracted data  on $F_2$ at low  $x$ with the GBW saturation model
and its modification to cover high values of $Q^2$. We find that the model stands the test of time and gives a good description of the data
with slightly modified parameters. All the essential elements of the model, especially the saturation scale, are retained.}

\keywords{Quantum Chromodynamics, deep inelastic scattering, parton distributions}

\begin{document}
\maketitle

\section{Introduction}
\label{sec:1}

{\comment The recently extracted data   \cite{Abt:2017nkc} on the proton structure function $F_2$ from the HERA  collider
summarizes  experimental effort  \cite{Aaron:2009aa,Abramowicz:2015mha}} done by the DESY collaborations H1 and ZEUS to maximally extend our knowledge
about the quark-gluon structure of the proton. The new kinematic region of small values of the Bjorken variable $x$, revealed by HERA,
corresponds to the high energy (or Regge) limit of QCD. In this limit, QCD enters the semihard perturbative domain where 
virtual photon "mass", $Q$, is much bigger than $\Lambda_{\rm QCD}$ and much smaller than invariant energy of the virtual photon-proton system, $W$,  {\it i.e.} when $x\approx Q^2/W^2\ll 1$. The first condition makes the running strong coupling, $\alpha_s(Q)$, small enough for perturbative calculations
to be applicable, leading to either the DGLAP \cite{Gribov:1972ri, Altarelli:1977zs, Dokshitzer:1977sg} or the BFKL \cite{Lipatov:1976zz,Kuraev:1976ge,Kuraev:1977fs,Balitsky:1978ic} evolution equations.  Both equations predict that the small $x$ structure of the proton is dominated
by a strongly rising  gluon density when $x\to 0$, which drives a similar rise  of the sea quark densities.

The problem of taming the growth of such densities, related to the question  about unitarity, was  pioneered in \cite{Gribov:1981ac,Gribov:1984tu}  and developed in \cite{Mueller:1985wy} in terms of nonlinear modifications of the known evolution equations, resulting from gluon recombination. A further refinement of the problem was proposed in  \cite{McLerran:1993ni,McLerran:1993ka,JalilianMarian:1996xn} as an effective field theory approach,
which  culminated in the formulation of 
the JIMWLK evolution equations  \cite{JalilianMarian:1997gr,Iancu:2000hn,Iancu:2001ad,Iancu:2003xm,Weigert:2005us} describing the over-populated gluonic state, called Color Glass Condensate (CGC).  An equivalent formulation was proposed in \cite{Balitsky:1995ub} in terms of the hierarchy of equations for Wilson line operators. This hierarchy simplifies in the limit of large number of colors $N_c$, which leads to the formulation of the Balitsky-Kovchegov (BK) equation for a dipole scattering amplitude in \cite{Kovchegov:1999yj,Kovchegov:1999ua}. 
The dipole amplitude is a key element in the  computation  of the small-$x$ DIS structure functions 
in the dipole picture in which the virtual photon splits into a $q\bar{q}$ dipole interacting with the proton through gluonic fields.
The dipole scattering amplitude obtained from the BK equation unitarizes the DIS structure functions by taming their power-like growth with $x\to 0$, which is a reflection of a similar growth of the gluon density, see \cite{GolecBiernat:2001if} for numerical studies of this effect. 

The solution to the BK equation provides theoretical justification of the $q\bar{q}$ dipole scattering amplitude (or a dipole cross section) in the GBW model proposed in \cite{GolecBiernat:1998js}  for phenomenological studies of the transition of   $F_2$ to small values of $Q^2$ and large values of $W^2$. With only three fitted  parameters, a good description of the DIS data was found. 
The dipole scattering amplitude  in the GBW model  vanishes for transverse dipole sizes  $r\to 0$   and saturates to a constant value for large $r$. These features make a connection to QCD for small $r$ (color transparency)  and to the idea of gluon saturation 
for large $r$ in which the number of gluons 
(or $q\bar{q}$ dipoles in large $N_c$ limit)  is tamed due to their recombination. The key element in the description  is an $x$-dependent saturation scale, 
$Q_s(x)$,  which sets the scale for dipole sizes.  With the proposed saturation scale, the dipole cross section saturates for smaller and smaller  
 dipole sizes with decreasing $x$. The GBW model was  updated  in \cite{Bartels:2002cj,GolecBiernat:2006ba} to improve the large $Q^2$ description of $F_2$
  by a modification of the small $r$ behaviour of the dipole cross section to include the DGLAP evolved gluon distribution. A similar in spirit parameterization of the dipole scattering amplitude, based on the BK equation solution, was proposed in \cite{Iancu:2003ge}.

{\comment  Since the publication  \cite{GolecBiernat:1998js}, numerous fits with the dipole picture of DIS were performed. 
Let us mention  the fits with the impact parameter
 dependent cross section \cite{Kowalski:2003hm,Rezaeian:2012ji,Rezaeian:2013tka} and the fits based on the solution to the BK equation
 \cite{Albacete:2009fh,Albacete:2010sy,Lappi:2013zma,Iancu:2015joa}.  
 Such studies are particularly important in  view of the analyses  \cite{Albacete:2012rx, Ball:2017otu}, 
 which show limitations  of the linear  DGLAP evolution for small $x$ and $Q^2$ values. 
 However, we should also mention  the analyses \cite{Ewerz:2007md,Ewerz:2011ph} on the limitations of the dipole models.
 }

 The main purpose of this presentation is  to answer the question whether 
 we still obtain a good description of
the new HERA data  \cite{Abt:2017nkc}   with the saturation model  \cite{GolecBiernat:1998js} and its modification  \cite{Bartels:2002cj,GolecBiernat:2006ba}. 
To this end, we preform new fits which update the original  values of the fit parameters. 
We include both charm and bottom quark contributions to $F_2$
(generated radiatively from gluons), which allows us to make predictions 
for the comparison with  data.
We also compute the longitudinal structure function $F_L$  to be compared with the data.

The paper is organized as follows. In Section \ref{sec:2} we present the results for the fits with the GBW and DGLAP improved models and make
a detailed comparison of them. In Section \ref{sec:3} we presents the comparison of the model results with the data on the proton structure functions, $F_2,F_2^{c,b}$ and $F_L$. Finally, we  summarize our findings in Conclusions.

\section{Fit results}
\label{sec:2}

We fit  the proton structure function $F_2$ computed from the HERA data \cite{Abt:2017nkc} on the $\gamma^* p$ cross section from the relation
\be
\sigma^{\gamma^*p}(x,Q^2)
=
\frac{4\pi^2\alpha_{em}}{Q^2}\,F_2(x,Q^2)\,,
\label{eq:1}
\ee
valid  in the small $x$ approximation. 
In all our fits we take the data with  $x\le 10^{-2}$ and minimal  value of $Q^2=0.045\,{\rm GeV}^2$.
 We consider both the light and heavy quarks (charm and bottom) in our calculations of the cross sections.

The basic theoretical formula for $F_2$ corresponds to the dipole picture of DIS at small~$x$ in which the virtual photon dissociates
 into a quark-antiquark pair  (a $q\bar{q}$ dipole) and subsequently interacts with the proton. Thus
\be
F_2=F_T+F_L=\frac{Q^2}{4\pi^2\alpha_{em}}\left(\sigma_T^{\gamma^*p}+\sigma_L^{\gamma^*p}\right)
\ee
where $T,L$ refer to virtual photon polarization, transverse and longitudinal, and
\be
\sigma_{T,L}^{\gamma^*p}=\sum_{f}\int d^2r\int_0^1dz\,|\Psi_{T,L}(r,z,Q^2,m_f)|^2\,\sigma_{\rm dip}(r,x)
\ee
where the sum over quark flavours $f$ is performed.
The photon wave function, $\Psi_{T,L}$, is known \cite{Bjorken:1970ah}, but the dipole cross sections is modelled with a few parameters which are fitted to the data.

Since the photon wave function depends on mass of the quarks in the $q\bar{q}$ dipole, we can consider contributions to the structure functions
from the individual quark flavour pairs
\be
F_{T,L}=F_{T,L}^{l}+F_{T,L}^{c}+F_{T,L}^{b}
\ee
where $F^l_{T,L}$ is the sum of  the contributions from the light quark pairs, $u\bar{u},d\bar{d}$ and $s\bar{s}$, while $F^{c}_{T,L}$ and $F^{b}_{T,L}$ 
 are the contributions
from the $c\bar{c}$ and $b\bar{b}$ pairs, respectively.
In such a case, we also modify the Bjorken variable 
$x$  in the dipole cross section
\be\label{eq:rescaledx}
x\to \bar{x}_f=x\left(1+\frac{4m_f^2}{{\red Q^2}}\right)=\frac{Q^2+4m_f^2}{Q^2+W^2}
\ee
where $W^2$ is an invariant energy squared of the $\gamma^*p$ system. The condition $\bar{x}_f\le 1$ accounts for the quark-pair production threshold.
{\comment However, to be in accord with the small $x$ approximation,  
we set the upper limit for the allowed values of $\bar{x}_f$  substituted in the dipole cross section
to $\bar{x}_f\le 0.1$.}  Above this value, the  charm and bottom structure functions are  equal to zero. {\red Nevertheless, almost all experimental points from HERA, 
 which we use in our analysis,  fulfill the bound  $\bar{x}_f\le 0.1$}.

\subsection{Fits with the GBW model}

The dipole cross section of the GBW model is given by \cite{GolecBiernat:1998js}
\be
\sigma_{\rm dip}(r,x) = \sigma_0\left(1-\eto^{-r^2Q_s^2(x)/4}\right)
\label{eq:2}
\ee
where the saturation scale $Q_s$ is defined as
\be
Q_s^2(x) = Q_0^2\, (x/x_0)^{-\lambda}
\label{eq:3}
\ee
with $Q_0^2=1\,{\rm GeV}^2$.  The above cross section has a property of geometric scaling \cite{Stasto:2000er},
\be
\sigma_{\rm dip}(r,x) = \sigma_{\rm dip}(r Q_s(x))\,,
\label{eq:8a}
\ee
\ie it becomes a function of a single variable, $rQ_s$,
for all values of $r$ and $x$. The three  parameters of the fits with the GBW model
are: $\sigma_0,\, \lambda$ and~$x_0$.

\begin{table}[t]
\begin{center}
\begin{tabular}{|c||c|c|c||c|c|c||c|}
\hline
Fit  &  $m_{l}$ &  $m_{c}$ & $m_b$ & $\sigma_0$[mb] &  $\lambda$   & $x_0/10^{-4}$  & $\chi^2/N_{\rm dof}$ \\
\hline\hline
 \cite{GolecBiernat:1998js}      & 0.14  & $-$   & $-$ &  23.02 & 0.288 & $3.04$  &  2.86\\
 \cite{GolecBiernat:1998js}     &  0.14  & 1.4 & $-$ & 29.12 & 0.277 & $0.41$  &    3.78\\   
  \hline 
 {0}     & 0.14  &  $-$ &  $-$  & $23.58\pm 0.28$& $0.270\pm 0.003$ & $2.24\pm 0.16$  &    400/219=1.83\\  
  {1}   &0.14  &  1.4 & $-$ & $27.32\pm 0.35$ & $0.248\pm 0.002$ & $0.42\pm 0.04 $ &  351/219=1.60 \\
 {2}  & 0.14  &  1.4 & $4.6$ & {$27.43\pm 0.35$} & {$0.248\pm 0.002$} & $0.40\pm 0.04$ &  352/219=1.61\\ 
 \hline
\end{tabular}
\end{center}
\caption{ Fit results to  the HERA data for  $Q^2\le 10\,{\rm GeV}^2$ with $N_p=222$ points, using the dipole cross section (\ref{eq:2}).  
Quark masses are given in units of GeV.  
$\chi^2/N_{\rm dof}$ computed with  the parameters from \cite{GolecBiernat:1998js} are given in the first two rows. 
 The parameters of Fit 2 are used for further analysis.
}
\label{table:gbw-fits}
\end{table}

\begin{figure}[b]
\begin{center}
\includegraphics[width=\textwidth]{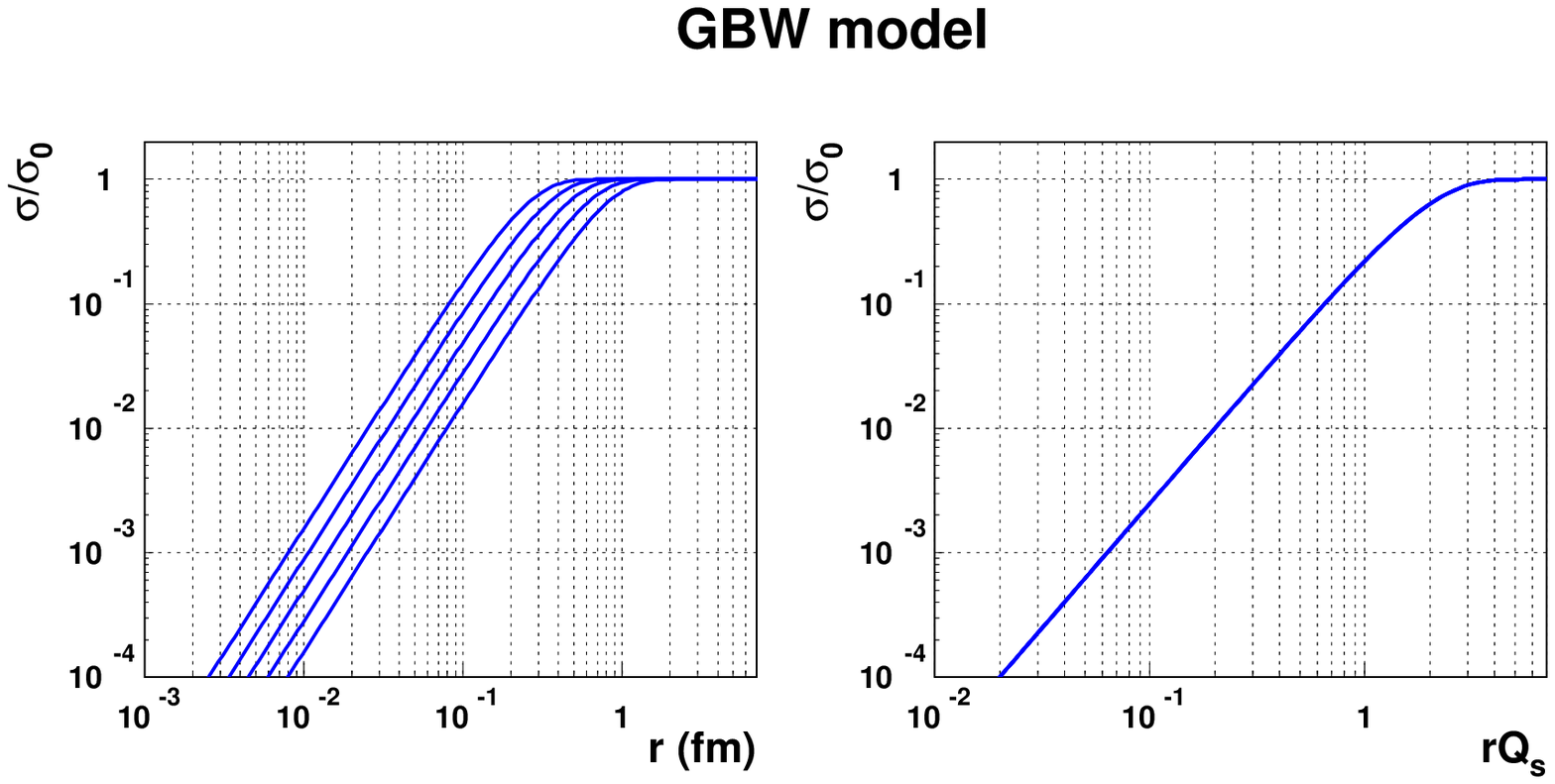}
\end{center}
\caption{Dipole cross section (\ref{eq:2})  with the parameters
from Fit 2 in Table~\ref{table:gbw-fits} as a function
 $r$ (left plot) and  $rQ_s$  (right plot) for $x=10^{-6},\ldots,10^{-2}$
(curves from left to right, respectively).  All the curves in the left plot merge into one line in the right plot due to geometric scaling (\ref{eq:8a}).
}
\label{fig:gbw-dcs}
\end{figure}

The GBW model does not incorporate QCD evolution in $Q^2$ in the small-$r$ part of the dipole cross section. As a consequence, the power $\lambda$, which governs change of the saturation scale with $x$, is
 independent of $Q^2$. Therefore, one does not expect the model to fare well for high
photon virtualities. In view of the above, we restrict data points used in the
fits to those with $Q^2 \le 10\,{\rm GeV}^2$ and $x\le 10^{-2}$, which leads to $N_p=222$ points.

The results of our fits with the new HERA data are summarized in Table~\ref{table:gbw-fits}.
In the first two rows  we show the value of $\chi^2/N_{\rm dof}$, where $N_{\rm dof}=N_p-(\#~{\rm of~parameters})$,
computed with the original parameters of the GBW model \cite{GolecBiernat:1998js}  (no fits were performed). 
Rather large values of $\chi^2/N_{\rm dof}$ suggest necessity of new fits. 

In the next three rows, Fits 0-2, we present the results
of the fits  which update the original parameters of the model.
{\comment The parabolic errors of  the fit parameters are given by MINOS from the MINUIT package \cite{James:1975dr,James:1994vla}, 
which we use in this paper to minimize $\chi^2$}.
We see a good fit quality, taking into account the data precision and 
a minimal number of fitted parameters. The new parameters are close to the original ones with smaller  
values of the parameter $\lambda$ by around $10\%$.
In addition, unlike the original fit results, adding charm into the analysis
improves the  data description.   The bottom quark contribution  is very small for  $Q^2\le 10\,{\rm GeV}^2$, nevertheless,   
we take the parameters of Fit 2 for further analysis to have a  full handle on heavy quarks.

In Figure~\ref{fig:gbw-dcs}, we show  the dipole cross section (\ref{eq:2}) with the parameters
from  Fit 2 computed  for $x=10^{-2},\ldots,10^{-6}$. These are 
the curves from  right to  left in the left plot  where the dipole  cross section is plotted as a function of the dipole size $r$. 
All the curves merge into one solid line  in the right plot where the dipole cross section is plotted
as a function of the  scaling variable $rQ_s$.  This is a reflection of  geometric scaling (\ref{eq:8a}) in the GBW model.

{\comment
Finally, we study the sensitivity of the fit quality to the choice of the maximal value of the photon virtuality, $Q^2_{\rm max}$,   in the data. 
The results of the fits with five flavours 
for the indicated  values of $Q^2_{\rm max}$ are shown  in Table \ref{table:gbw-fits-q2max}. 
We refrain from quoting  the fit parameter errors, whose numerical values are very close to those from Fit  2 in Table  \ref{table:gbw-fits}.
We see that the choice $Q^2_{\rm max}=10\,{\rm GeV^2}$ is indeed optimal, although  extending the applicability of the GBW model up to 
$Q^2_{\rm max}=20\,{\rm GeV^2}$
is still acceptable.
}

\begin{table}[t]
\begin{center}
\begin{tabular}{|c|c||c|c|c||c|}
\hline
  $Q^2_{\rm max} [{\rm GeV}^2]$ &~~~$N_p$~~~&  $\sigma_0$[mb] &  $\lambda$   & $x_0/10^{-4}$  & $\chi^2/N_{\rm dof}$ \\
\hline\hline
   5 &   181  &$28.18$ & $0.237$ & $0.31$ &  292/178=1.64\\ 
  10 &   222  &$27.43$ & $0.248$ & $0.40$ &  352/219=1.61\\ 
  20 &   264 &$26.60$ & $0.259$ & $0.53$ &  430/261=1.65\\ 
  50 &   318  &$25.21$ & $0.281$ & $0.80$ &  764/315=2.43\\   
  \hline
\end{tabular}
\end{center}
\caption{ {\comment Sensitivity of the quality of Fit 2 from Table \ref{table:gbw-fits} to the choice of $Q^2_{\rm max}$ in the data.
$N_p$ is the number of experimental points selected by  $Q^2_{\rm max}$.}
}
\label{table:gbw-fits-q2max}
\end{table}

Summarising, the GBW model gives a good description of the HERA data for small and moderate values of
$Q^2$, accounting for  the  transition of the DIS structure functions to small values of $Q^2\le 1\,{\rm GeV}^2$ This is achieved with the idea of parton saturation and only three fitted parameters. A graphical illustration of the agreement of the model with the data will be presented in Section \ref{sec:3}.

\subsection{Fits with the DGLAP improved saturation model}

The DGLAP improved saturation model \cite{Bartels:2002cj,GolecBiernat:2006ba}
implements the dipole cross section given by
\be
\label{eq:33}
\sigma_{\rm dip}(r,x)=
\sigma_0\left\{1-\exp\left(-\frac{\pi^2r^2\,\alpha_s(\mu^2)\,xg(x,\mu^2)}{3\sigma_0}\right)\right\}\,,
\ee
where $g(x,\mu^2)$ is the gluon distributions taken at the scale
\be
\label{eq:scale-dglap-ver1}
\mu^2=\frac{C}{r^2}+\mu_0^2\,.
\ee
The gluon distribution is evolved with the DGLAP evolution  equations truncated to the gluonic sector, 
\be
\frac{\partial g(x,\mu^2)}{\partial \ln \mu^2}=\frac{\alpha_s(\mu^2)}{2\pi}\int_x^1\frac{dz}{z}\,P_{gg}(x)\,g(x/z,\mu^2)\,,
\ee
from the initial condition
\be
\label{eq:5}
xg(x,Q_0^2)=A_g\,x^{-\lambda_g}\,(1-x)^{5.6}\,,
\ee 
taken at the scale $Q_0=1\,{\rm GeV}$. The choice of the power $5.6$, which regulates the large-$x$ behaviour, 
is motivated by global fits to DIS data with the LO DGLAP equations, see \cite{Bartels:2002cj,GolecBiernat:2006ba} for more details.
The splitting function $P_{gg}$ contains real and virtual terms with the number
of active quark flavours $n_f$ in the latter one
\be
P_{qq}(z)=2N_c\left(\frac{z}{(1-z)_+} +\frac{1-z}{z}+z(1-z)\right)+\delta(1-z)\frac{11C_A-4n_fT_R}{6}
\ee
with $C_A=N_c=3$ and $T_R=1/2$. In the leading order strong coupling constant we set $\Lambda_{\rm QCD}=300\,{\rm MeV}$.
Thus, the model has  five parameters to fit: $\sigma_0,  A_g, \lambda_g, C$ and  $\mu_0^2$.

\begin{table}[t]
\begin{center}
\begin{tabular}{|c||c||c|c|c|c|c||c|}
\hline
Fit   &   $m_b$ & $\sigma_0$[mb]  & $A_{g}$ & $\lambda_{g}$    & $C$ &  $\mu_0^2 [{\rm GeV}^2]$ &  
 $\chi^2/N_{\rm dof}$ \\
 \hline\hline
\cite{GolecBiernat:2006ba}      & $-$     & {22.40} &$1.35$  &$0.079$ &  0.38 &  $1.73$ &   2.02\\  
\cite{GolecBiernat:2006ba}      & $4.6$ & {22.70} &$1.23$  &$0.080$ &  0.35 &  $1.60$ &   2.43\\  
 \hline
{1}    &  $-$ & $22.60\pm 0.26$  & $1.18\pm 0.15$  & $0.11\pm 0.03$ &  $0.29\pm 0.05$ & $1.85\pm 0.20$ & 536/382=1.40 \\ 
{2}    & 4.6 & $22.93\pm 0.27$  &$1.07\pm 0.13$  & $0.11\pm 0.03$ &  $0.27\pm 0.04$ & $1.74\pm 0.16$ & 578/382=1.51\\ 
\hline
\end{tabular}
\end{center}
\caption{ {\comment 
The  results of the fits to the HERA data for  $Q^2\le 650\,{\rm GeV}^2$ with $N_p=387$ points, using the dipole cross section (\ref{eq:33}) with the scale  (\ref{eq:scale-dglap-ver2}). 
The quark masses $m_l=0$ and $m_c=1.3\,{\rm GeV}$.
 $\chi^2/N_{\rm dof}$ computed with  the parameters from \cite{GolecBiernat:2006ba} and the scale (\ref{eq:scale-dglap-ver1}) are given in the first two rows. }
The parameters of Fit 2 are used for further analysis.
}
\label{table:dglap-fits}
\end{table}

\begin{figure}[t]
\begin{center}
\includegraphics[width=\textwidth]{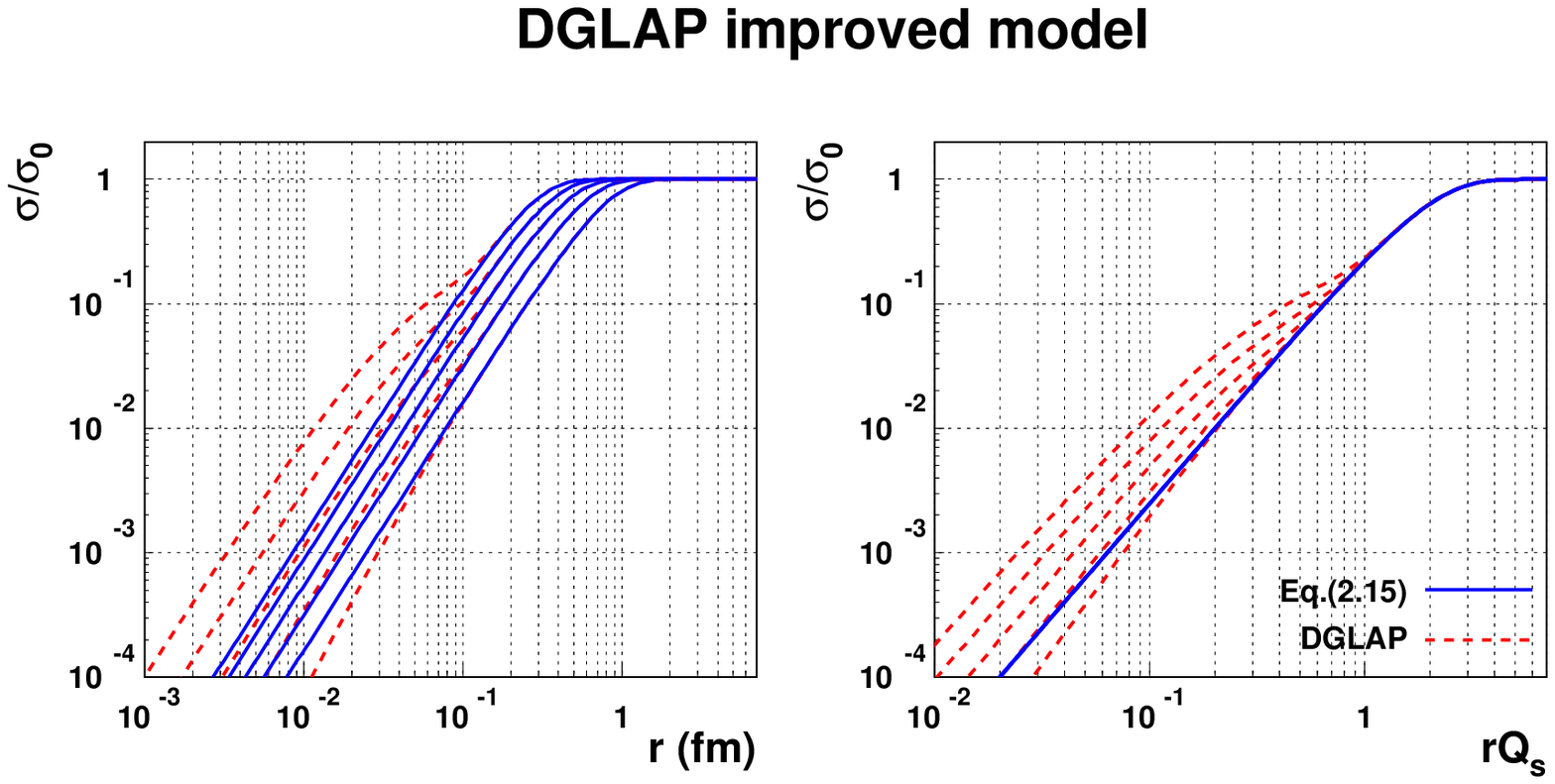}
\end{center}
\vskip -5mm
\caption{Dipole cross section (\ref{eq:33}) for the scale (\ref{eq:scale-dglap-ver2})  with the parameters
from Fit 2 in Table~\ref{table:dglap-fits}  (dashed lines)  as a function  $r$ (left plot) and $rQ_s$  (right plot) for
$x=10^{-6},\ldots,10^{-2}$ (curves from  left to right, respectively). The solid lines, 
corresponding to Eq.~(\ref{eq:7}),  merge into one solid line in the right plot
due to geometric scaling with the  saturation scale (\ref{eq:8}).  For $rQ_s\ge 1$, also the dashed curves merge  due to geometric scaling in
the dipole cross section (\ref{eq:33})
in this region.}
\label{fig:fig2}
\end{figure}

The dipole cross section (\ref{eq:33}) has the property of colour transparency and 
 tends to the perturbative  QCD result in the limit $r\to 0$. Indeed, for small dipoles, the scale $\mu^2\approx C/r^2 $ and the dipole cross section
is proportional to $r^2$ with the logarithmic  modifications due to the scale
dependence of the gluon distribution \cite{Frankfurt:1996ri}, 
\be
\label{eq:6}
\sigma_\text{dip}\approx \frac{\pi^2}{3}r^2\alpha_s(C/r^2)\,xg(x,C/r^2)\,.
\ee
These additional logarithms  allow  better fits to the data for large values of  $Q^2$. 
In the limit of  large dipoles  $\mu^2\approx \mu_0^2$, which
leads to 
\be
\label{eq:7}
\sigma_{\rm dip}\approx \sigma_0\left\{1-\exp\left(-\frac{\pi^2r^2\,\alpha_s(\mu_0^2)\,xg(x,\mu_0^2)}{3\sigma_0}\right)\right\}.
\ee
Thus, at large $r$, we find the GBW form of the dipole cross section with the saturation scale
\be
\label{eq:8}
Q_s^2(x)= \frac{4\pi^2}{3\sigma_0}\,\alpha_s(\mu_0^2)\,xg(x,\mu_0^2)\,,
\ee
with the  $x$ dependence given by the gluon distribution taken at the scale $\mu_0^2$.
Geometric scaling  is strictly valid for (\ref{eq:7}), which is not the case for
small dipoles when an additional $r$ dependence  is introduced in the dipole cross section (\ref{eq:33}) through
the scale (\ref{eq:scale-dglap-ver1}).

The above features of the dipole cross section can also be obtained
 for a slightly different  choice of the scale $\mu$, 
 \be
\label{eq:scale-dglap-ver2}
  \mu^2=\frac{\mu_0^2}{1-\exp(-\mu_0^2\,r^2/C)}\,,
\ee
which interpolates smoothly  between the $C/r^2$ behaviour for small $r$ and the constant behaviour, $\mu^2=\mu_0^2$,
 for $r\to\infty$. The fit quality is  better for such a choice, thus in the forthcoming,
we present the results of the fits with the above scale.

The results of the fits are presented in Table\,\ref{table:dglap-fits}. The parabolic errors of the fit parameters are given by MINOS from the MINUIT package.
We no longer restrict the data  to low $Q^2$ values since  the DGLAP evolution is incorporated in this model. 
Now, the data set contains $N_p=387$ points with $Q^2\le 650\,{\rm GeV}^2$ and $x\le 10^{-2}$.
 We also no longer consider the case with light quarks only since the heavy quark contribution to $F_2$ (especially from charm) becomes important for large values of $Q^2$. In addition, we use the PDG world average value of the charm quark mass, 
 $m_c=1.275\pm 0.025\,{\rm GeV}\approx 1.3\,{\rm GeV}$ \cite{Beringer:1900zz}.
 
In the first two rows of Table\,\ref{table:dglap-fits}, we show $\chi^2/N_{\rm dof}$ computed with the original parameters of the DGLAP improved
model \cite{GolecBiernat:2006ba} and the prescription (\ref{eq:scale-dglap-ver1}) for the scale $\mu$  (no fits were performed).
In the next two rows, Fits 1-2, we present the fit results to the new HERA data with the scale (\ref{eq:scale-dglap-ver2}). 
We find  much better description of the data after refitting the parameters.
The nonzero light quark mass (equal to $140\,{\rm MeV}$ in the GBW model) deteriorates the fit quality,  thus we set it to zero,
$m_l=0$. For the rest of the presented analysis, we use the parameters from Fit 2 in Table 2 with the charm and bottom contributions included.

\begin{figure}[t]
\begin{center}
\includegraphics[width=\textwidth]{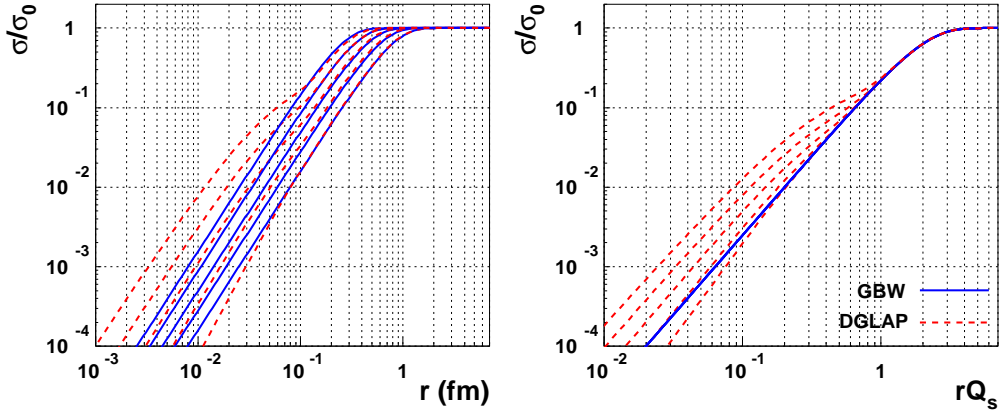}
\end{center}
\vskip -5mm
\caption{The normalized-to-one dipole cross sections (\ref{eq:2})  (solid
lines) and (\ref{eq:33}) (dashed lines) with the  parameters from Fit 2 in Table~\ref{table:gbw-fits} and
\ref{table:dglap-fits}, respectively, for $x=10^{-6},\ldots,10^{-2}$ (curves from left to right, respectively)
Geometric scaling for the variable $rQ_s$ 
is clearly visible in the right plot.}
\label{fig:fig3}
\end{figure}

The dipole cross section (\ref{eq:33})   from Fit 2  is shown in Figure~\ref{fig:fig2} (left)
as the dashed red lines corresponding to  different values of $x=10^{-2},\ldots,10^{-6}$  (from right to left). 
The blue solid lines show to the dipole cross section (\ref{eq:7}),  plotted in the whole range of $r$. Geometric scaling in this dipole cross section is clearly visible  as the single blue line in the right plot.  Since Eq.\,(\ref{eq:7}) is a limiting form  the dipole cross section (\ref{eq:33}), it is not surprising that geometric scaling is also visible for the dashed curves in the region $rQ_s\ge 1$, together with its violation for $rQ_s<1$.

Summarising,  the DGLAP improved model allows to extend the GBW saturation model to large values of $Q^2$, giving a good description of the HERA data up to $Q^2=650\,{\rm GeV}^2$. This is achieved by the modification of the dipole cross section for small dipole sizes to match the perturbative QCD result in this region. The essential elements of the GBW model, the saturation scale and geometric scaling, are retained in the DGLAP improved dipole cross section in the region $rQ_s\ge 1$, which  is mostly responsible for the transition of $F_2$ to small $Q^2$ values.

\subsection{Fit comparison}

\begin{figure}[t]
\begin{center}
\includegraphics[width=0.52\textwidth]{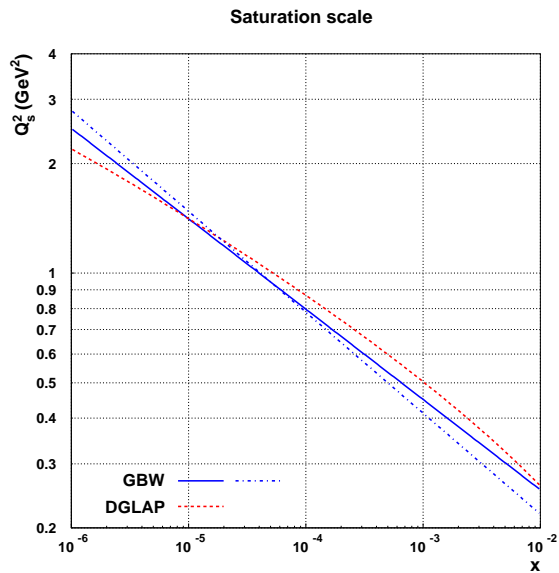}
\end{center}
\vskip -5mm
\caption{
The saturation scale  in the GBW model (solid line), Eq.\,(\ref{eq:3}),   and DGLAP improved model  (dashed line), Eq.\,(\ref{eq:8}), with the parameters from Fit 2 in Table~\ref{table:gbw-fits} and \ref{table:dglap-fits}, respectively. The saturation scale from the original GBW model  with charm \cite{GolecBiernat:1998js} is  shown as the dash-dotted line. 
}
\label{fig:fig4} 
\end{figure}

In Figure~\ref{fig:fig3} we compare the normalised-to-one dipole cross sections
from the GBW (blue solid lines) and  DGLAP improved (red dashed lines) models with the parameters  from Fit~2
in Table~\ref{table:gbw-fits} and \ref{table:dglap-fits}, respectively. We also use the appropriate saturation scales, given by Eqs.~(\ref{eq:3}) and (\ref{eq:8}),
for the scaling variable, $\rhat\equiv r Q_s$.   We see in the  left plot that for large values of $r$ the two functions overlap while they differ in 
the small-$r$ region where the running of the gluon distribution starts to play a significant role. This is clearly seen in the right plot
where geometric scaling holds for the DGLAP improved model curve for  the scaling variable $\rhat\ge 1$
and in the whole region   for the GBW model curve. The two model functions overlap for $ \rhat\ge1$ due to their universal form, 
$\sigma_{\rm dip}\sim 1-\exp(-\rhat^2/4)$, in this region.

Notice also that the leftmost dashed curve  of the DGLAP model in the left plot in Figure~\ref{fig:fig3}, 
corresponding to  $x=10^{-2}$, lies below the analogous GBW curve for small $r$. This is due to the suppression  term $(1-x)^{\alpha}$ present in the gluon distribution. 
Such a term is missing  in the GBW dipole cross section, where the saturation scale is always proportional to $x^{-\lambda}$.

It is also interesting to compare the saturation scales from the fits.  They are shown in Figure~\ref{fig:fig4} in the two analysed models 
as the blue solid (GBW model) and red dashed (DGLAP improved model) lines. For the reference, we also plot the saturation scale from 
the original GBW model \cite{GolecBiernat:1998js} with charm  as the blue dot-dashed line. We see that all lines lie close to each other, although their slope is slightly different. 
As a result, at $x=10^{-6}$ the  saturation scale $Q_s^2\approx 2-3\,{\rm GeV}^2$.

\begin{figure}[t]
\begin{center}
\includegraphics[width=0.49\textwidth]{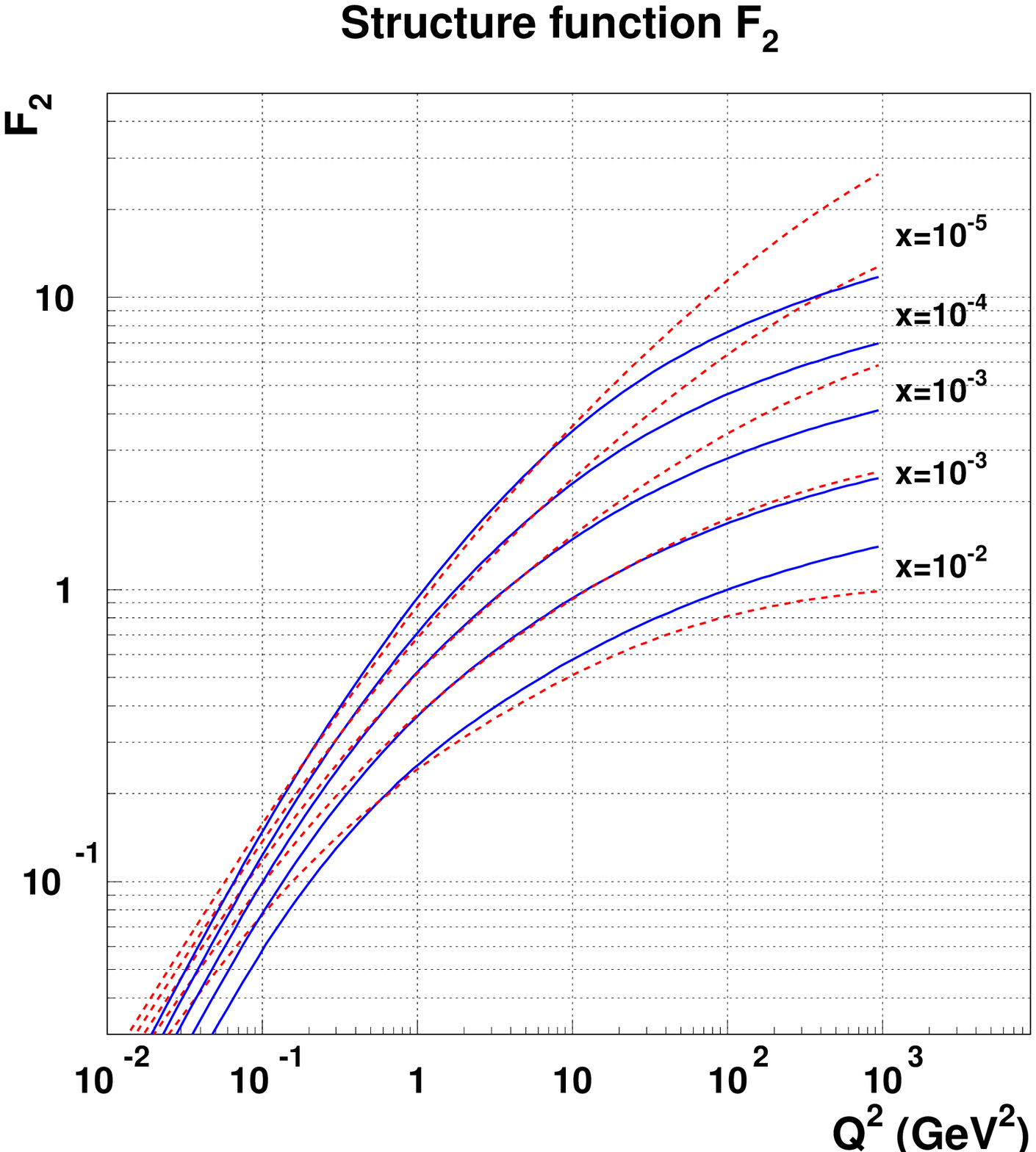}
\includegraphics[width=0.49\textwidth]{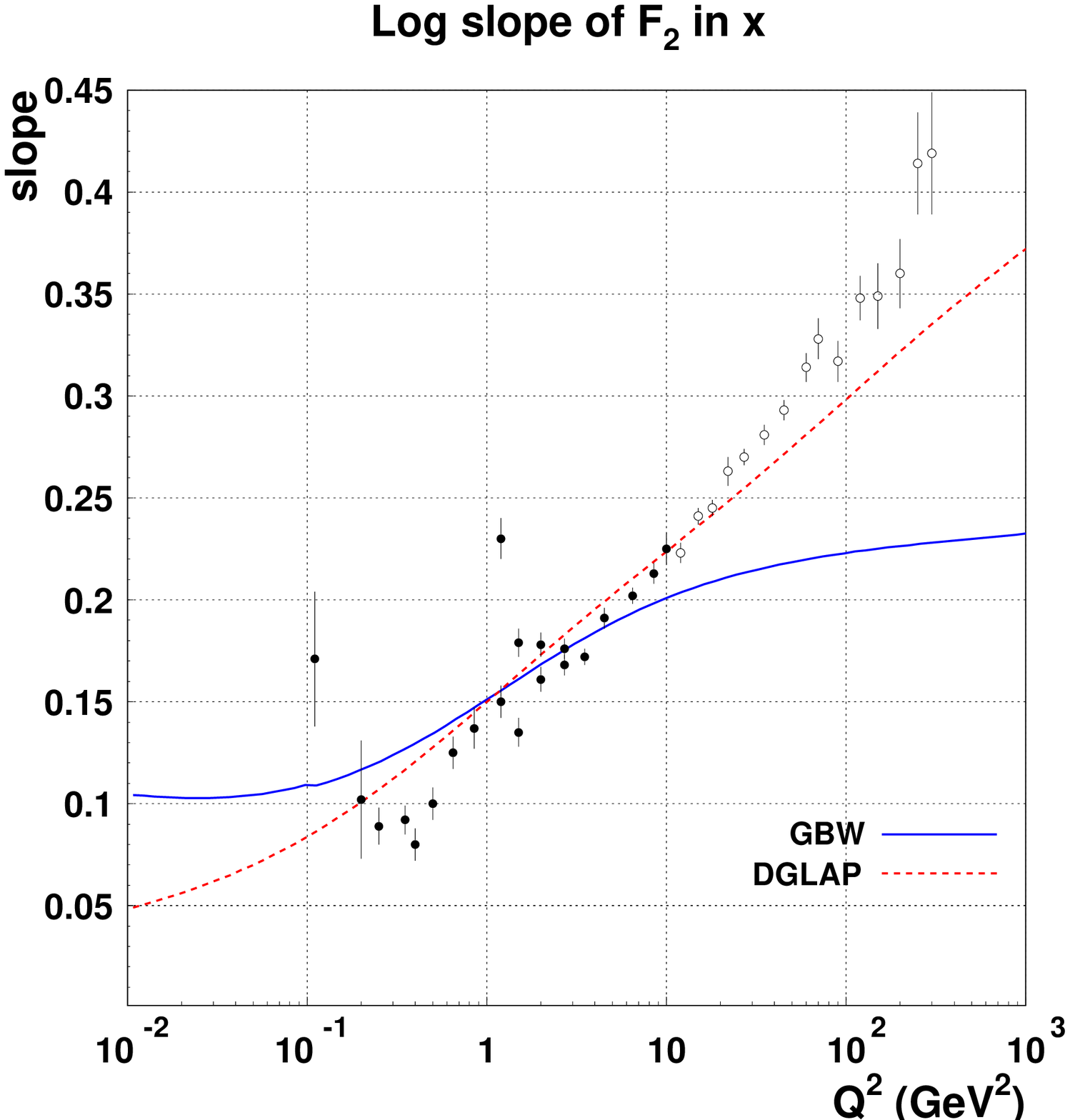}
\end{center}
\caption{Left plot: $F_2$ for fixed values of $x$ as a function of $Q^2$ for the GBW (solid line) and DGLAP improved (dashed lines) models. 
The logarithmic slope in $x$ of $F_2$ as a function of  $Q^2$ is shown in the right plot for the same models. 
The experimental points are from Ref.~\cite{Abt:2017nkc}.
}
\label{fig:fig3a}
\end{figure}

The structure function $F_2$ computed with the two dipole cross sections as a function of $Q^2$ for fixed values of $x$ is shown in the left plot in Figure~\ref{fig:fig3a}. We see that for $Q^2> 10\,{\rm GeV}^2$,  $F_2$ computed in the DGLAP improved model
rises stronger for small values of $x$. This is an effect of the DGLAP evolution of the gluon distribution in the dipole cross section. In the right plot, we show the logarithmic slope of $F_2$ in $x$ as function of $Q^2$, found from the approximate relation
\be
F_2\sim x^{-\lambda(Q)}\,,
\ee
valid for $x\to 0$. In the GBW model (blue solid line), the slope $\lambda(Q)$ tends to a constant value 
$\lambda$ in the saturation scale (\ref{eq:3}) for increasing values of $Q^2$, while in the DGLAP improved model (red dashed line) $\lambda(Q)$ 
strongly rises due  to the gluon evolution in the double logarithmic limit.  
On the other hand, for small $Q^2$ values, the GBW curve approaches  a constant
value close to the soft pomeron intercept, $\alpha_\funp(0)=1+\lambda\approx 1.1$. 
{\comment
The value of the slope depends on the light quarks mass $m_l$.  For example, computing
the slope  in the GBW model along the constant values of $x\in [10^{-5}, 10^{-3}]$ with  $m_l=140\,{\rm MeV}$, 
we find the value at $Q^2=10^{-2}\,{\rm GeV}^2$ shown on the plot,  $\lambda=0.1\pm 0.01$.  For $m_l=0$ we obtain $\lambda=0.045\pm 0.005$ and for $m_l=280\,{\rm MeV}$ we have $\lambda=0.15\pm 0.02$. In the latter case, however,
the fit quality is rather bad, $\chi^2/N_{\rm dof}=2.78$.
}

\section{Comparison to data}
\label{sec:3}

\begin{figure}[t]
\begin{center}
\includegraphics[width=0.49\textwidth]{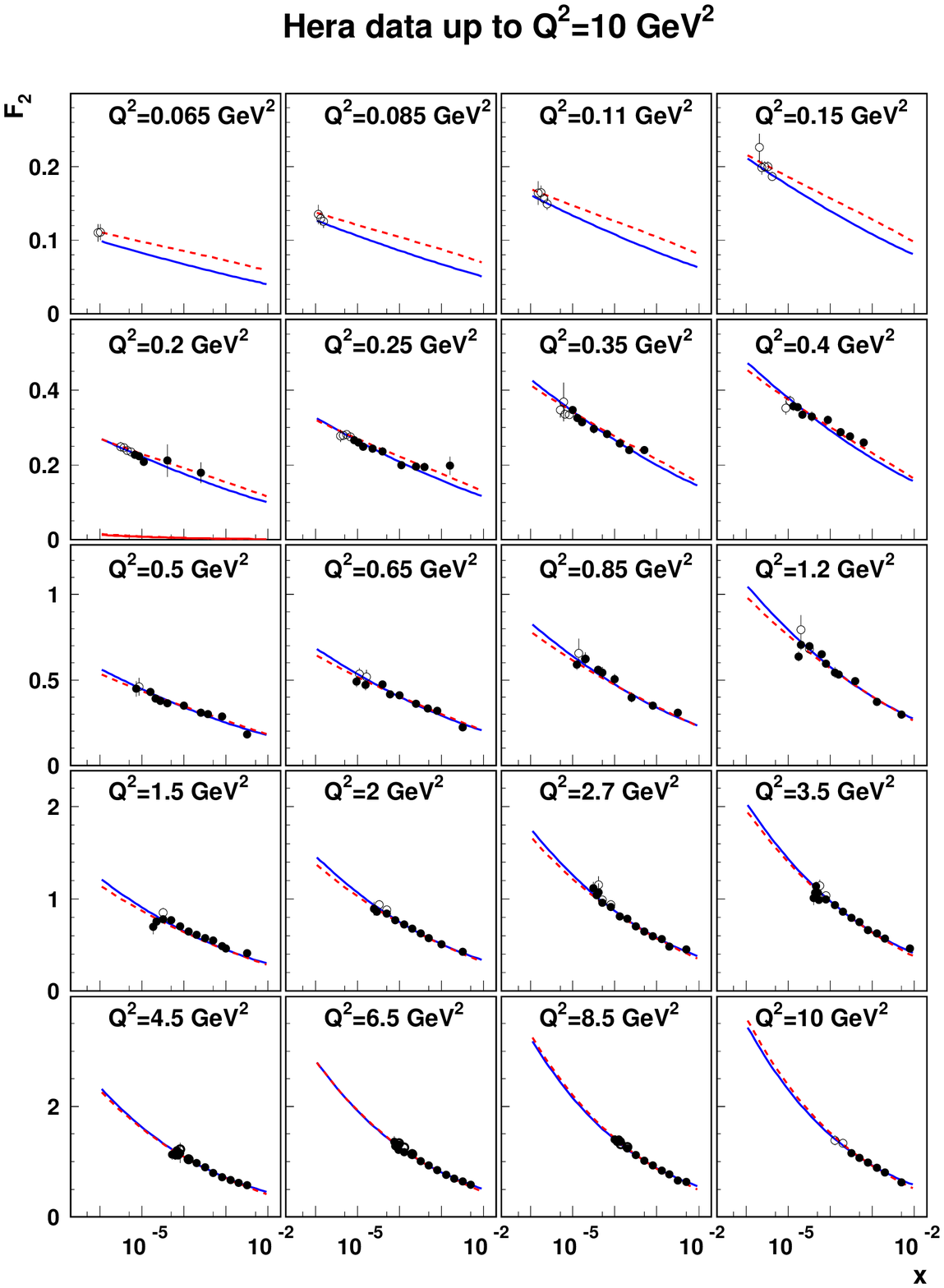}
\includegraphics[width=0.49\textwidth]{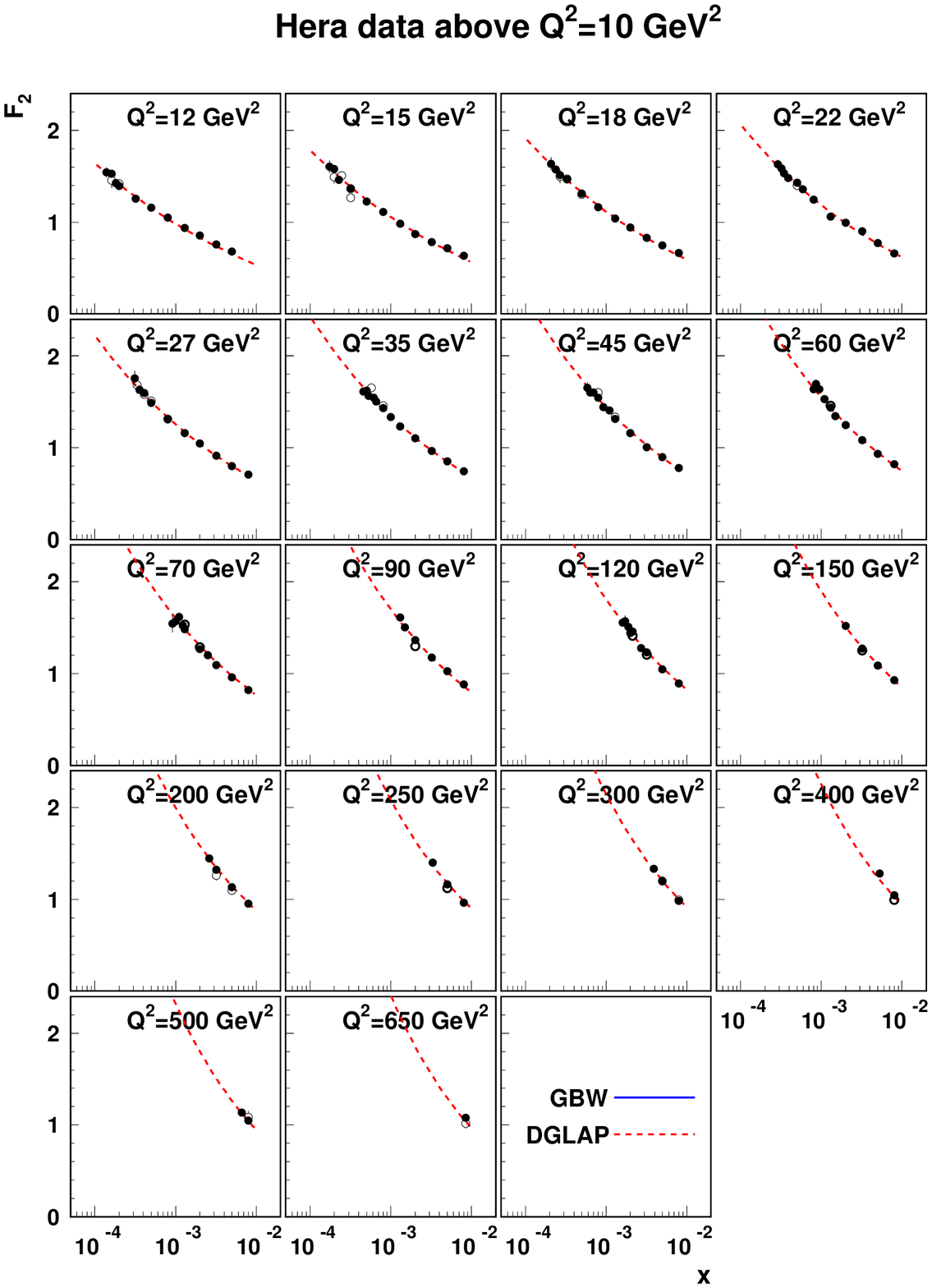}
\end{center}
\caption{Comparison of the HERA data  from \cite{Abt:2017nkc}  up to $Q^2=10\,{\rm GeV}^2$ (left plot) and above  (right plot) 
with the  results from   the GBW model (solid lines) and DGLAP improved model (dashed lines)  with the parameters 
of Fit 2 in Table~\ref{table:gbw-fits} 
and \ref{table:dglap-fits}, respectively. The GBW  model lines are shown only in the region where the model was fitted.
}
\label{fig:fig5}
\end{figure}

\begin{figure}[t]
\begin{center}
\includegraphics[width=0.49\textwidth]{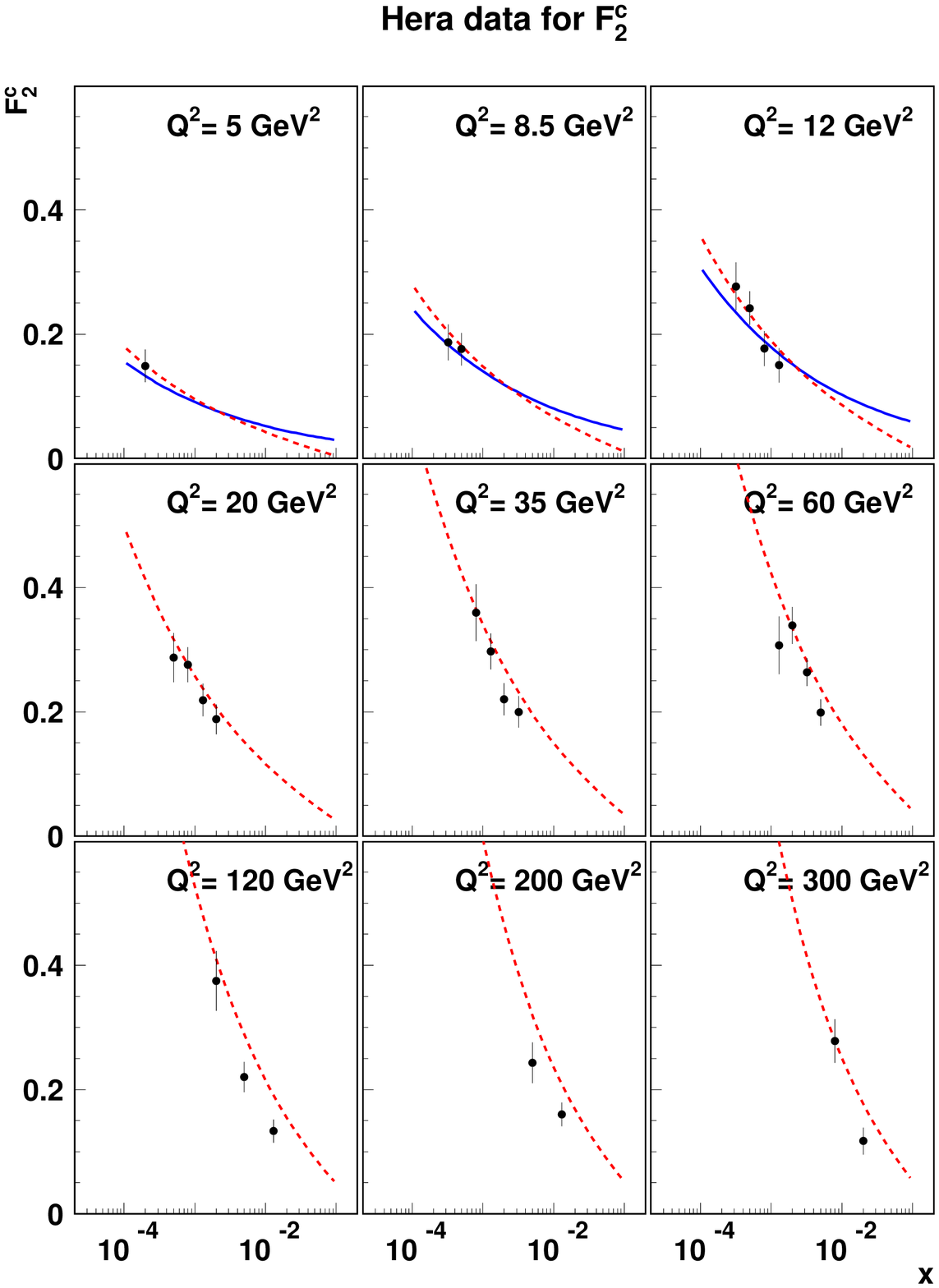}
\includegraphics[width=0.49\textwidth]{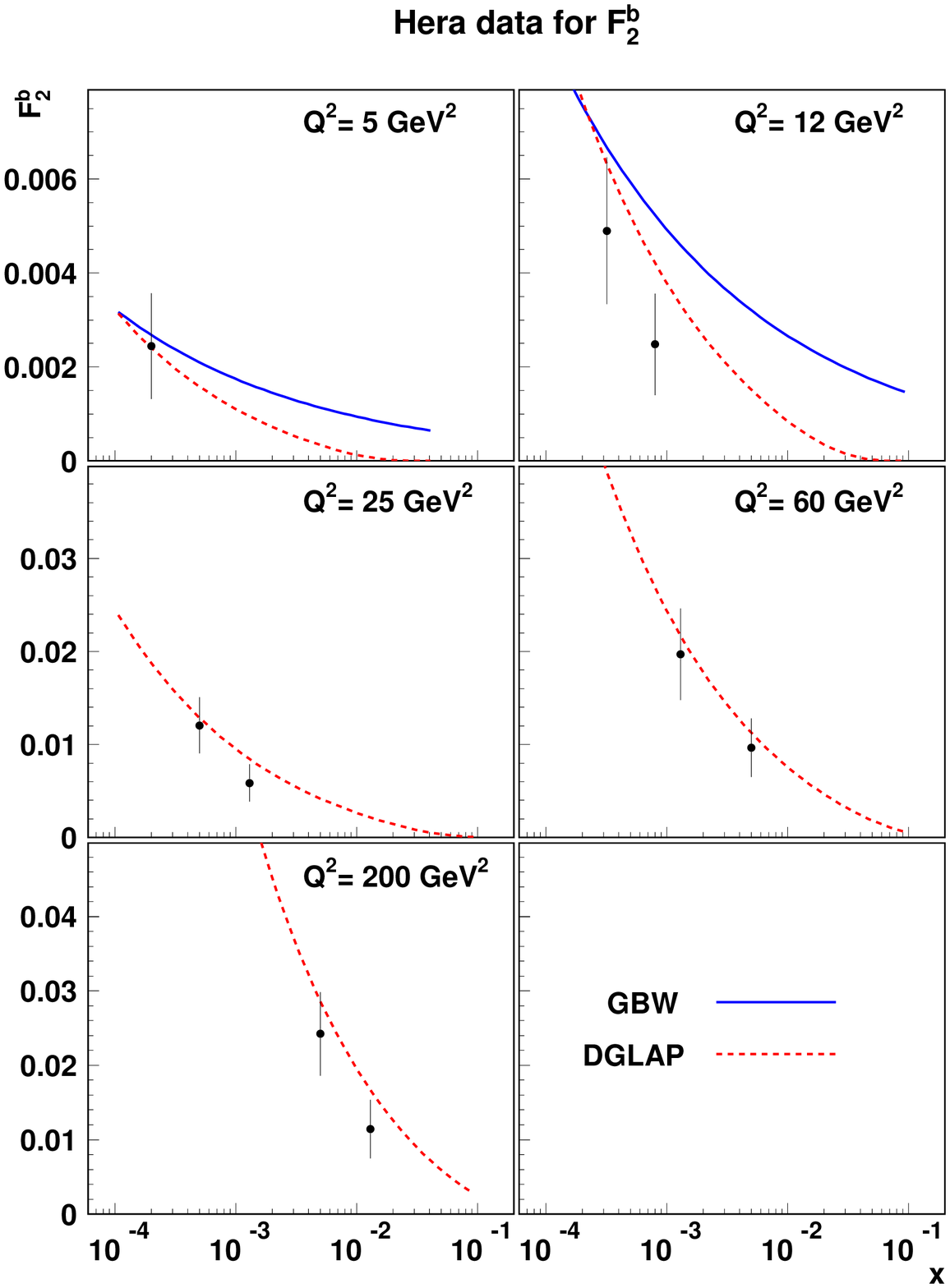}
\end{center}
\caption{Comparison of the HERA data from \cite{Aaron:2009af} for the charm $F_2^c$ (left panel) and bottom $F_2^b$ (right panel)  structure functions
with the  results from  the GBW model (solid lines) and DGLAP improved model (dashed lines) with the parameters of  Fit 2 in Table~\ref{table:gbw-fits} 
and \ref{table:dglap-fits}, respectively. The GBW  model lines are only shown in the region where the model was fitted and slightly above.
}
\label{fig:fig6}
\end{figure}

\begin{figure}[t]
\begin{center}
\includegraphics[width=0.7\textwidth]{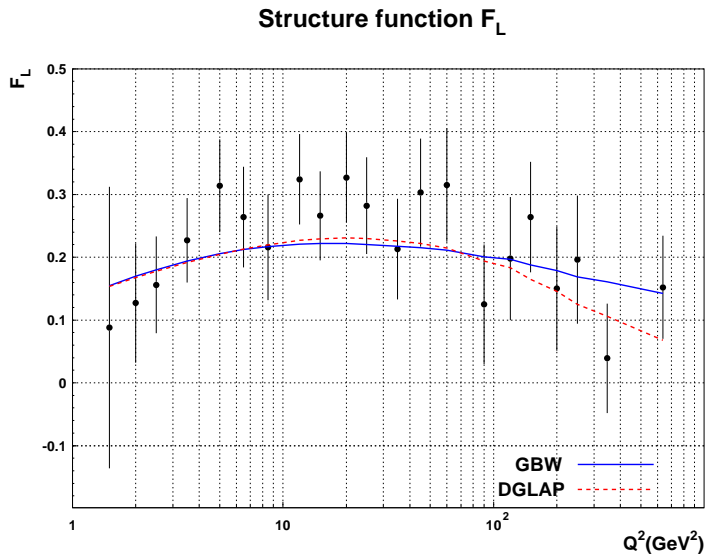}
\end{center}
\vskip -5mm
\caption{The longitudinal structure function $F_L$ in  the GBW model (solid line) and DGLAP improved model (dashed line) with the parameters  from Fit 2 in Table~\ref{table:gbw-fits} and \ref{table:dglap-fits}, respectively, against the HERA data from \cite{Andreev:2013vha}.}
\label{fig:fig7}
\end{figure}

In Figures \ref{fig:fig5}-\ref{fig:fig7},  we present the comparison of the predictions from the two models  discussed above with the newest HERA data. 

Finally, in Figure~\ref{fig:fig5}, we show
the structure function $F_2$ as a  function of $x$ for the indicated values of $Q^2$. 
We show separately the comparison with the data in the region up to $Q^2\le 10\,{\rm GeV}^2$ (left plot) and above this value (right plot).
This is to  indicate that the GBW model fits were only performed in the first region. 
 We see that good values of $\chi^2/N_{\rm dof}$ of the fits are reflected in a very good agreement with the data. 
 In the region of small and moderate values of $Q^2$ (left plot), the two models give similar results. Only for the lowest values of $Q^2$, the results slightly differ 
 because of nonzero light quark mass in the GBW model. For lager values of $Q^2$ (right plot),  the DGLAP improved model curves  also agree very well with the data.

A similar comparison, now for the charm and bottom contributions, $F_2^{c,b}$ respectively,   to the structure function $F_2$, is shown in Figure~\ref{fig:fig6} for the HERA data \cite{Aaron:2009af} with $Q^2\ge 5\,{\rm GeV}^2$. {\comment Note that the combined data from ZEUS and H1 for the reduced
cross section  for charm production, $\sigma^{c\bar{c}}_{\rm red}$, was published in \cite{Abramowicz:1900rp}. We stick, however, to the comparison with the structure function data.}
The model values of $F_2^{c,b}$ are genuine predictions since these contributions are not separately fitted to the experimental data on 
$F_2^{c,b}$. Instead, they are  determined  from the fits to the data on the total structure function $F_2$. We find good agreement with  the data for both  the DGLAP improved model (red dashed curves) and  the GBW model in the region of  $Q^2\le 10\,{\rm GeV}^2$ (blue solid lines). 

Finally, in Figure~\ref{fig:fig7} we show the comparison with the HERA data on the longitudinal structure function $F_L$ \cite{Andreev:2013vha}. 
Both models give  predictions which 
are in the right ballpark. We have to remember  that the GBW model was fitted to the data with $Q^2_{\rm max}=10\,{\rm GeV}^2$, and the curves
above this value are only  extrapolations. The difference between the two set of curves can be attributed to the lack of the suppression term $(1-x)^\alpha$
in the GBW model structure functions. 
{\comment This  term is  relevant for  the values of $x\sim 10^{-2}$ and the three data points with the largest
$Q^2$ in Figure~\ref{fig:fig7} have such values.}
The experimental precision  of the data prevents us from drawing  more precise conclusions.

\section{Conclusions}
\label{sec:4}

{\comment The new data  \cite{Abt:2017nkc}  on the proton structure function $F_2$, extracted from the HERA measurements  
\cite{Aaron:2009aa,Abramowicz:2015mha},}  
prompted us to address the question how the saturation model of DIS \cite{GolecBiernat:1998js,Bartels:2002cj,GolecBiernat:2006ba} describes this data for small values of the Bjorken variable, $x~\le 10^{-2}$, and $Q^2$ in the range between $0.065\,{\rm GeV}^2$ and $650\,{\rm GeV}^2$.

In the first part of our analysis, we only considered the data on $F_2$ for  low and moderate values of $Q^2\le 10\,{\rm GeV}^2$
to fit  three parameters of the GBW model  \cite{GolecBiernat:1998js}, which was originally devised to describe the transition of $F_2$ to small values of $Q^2< 1\,{\rm GeV}^2$. We found a good fit quality with the charm and bottom quark contributions to $F_2$ included in the fits. The refitted
parameters are close to the original values from \cite{GolecBiernat:1998js} with around $10\,\%$ decrease of the power $\lambda$ in the saturation scale (\ref{eq:3}),  see Table~1. However, 
as shown in Figure~\ref{fig:fig4}, the position of the saturation line on the $(x,Q^2)$-plane has not changed significantly. Therefore, we conclude
that the GBW model  still
provides an economical description of the data on the DIS structure functions, $F_2,F_L$ and $F_2^{c,b}$,  for the  values of
 $Q^2\le 10\,{\rm GeV}^2$.

In the second part of our study, we used the DGLAP improved model of the dipole cross section with saturation, which allows one to perform a good
quality fits to the new data with $Q^2\le 650\,{\rm GeV}^2$. In this model, the small dipole size part of the dipole cross section is modified
by the presence of the gluon distribution evolved with the DGLAP evolution equation. We proposed a new prescription 
(\ref{eq:scale-dglap-ver2}) for the scale of the gluon distribution which gives  better results than the original scale in 
\cite{Bartels:2002cj,GolecBiernat:2006ba}. As a result,  we obtain a dipole cross section which   for large dipoles
retains the features of the GBW model with the saturation scale given effectively by Eq.~(\ref{eq:8}), see Figure \ref{fig:fig3}. This scale is close to the scale from
the GBW model, see the red dashed line in Figure~\ref{fig:fig4}. Thus, we showed the robustness of the analysed saturation models, verified by the agreement of  with the data. 

In conclusion, the saturation model has stood the test of time and is still  a valuable contribution to our understanding of  DIS at small $x$\footnote{
\comment The code with the parametrization of the dipole cross sections and structure functions from Fits 2 in Table \ref{table:gbw-fits} and \ref{table:dglap-fits}  can be downloaded from \url{http://nz42.ifj.edu.pl/~sapeta/gbw2.0.html}.
}

\acknowledgments
This work was supported by the National Science Center grant No. 2015/17/B/ST2/01838,   
 and by the Center for Innovation and Transfer of Natural Sciences and Engineering Knowledge in Rzesz\'ow.


\begin{thebibliography}{10}

\bibitem{Abt:2017nkc}
I.~Abt, A.~M. Cooper-Sarkar, B.~Foster, V.~Myronenko, K.~Wichmann and M.~Wing,
  \emph{{Investigation into the limits of perturbation theory at low $Q^2$
  using HERA deep inelastic scattering data}},
  \href{http://dx.doi.org/10.1103/PhysRevD.96.014001}{\emph{Phys. Rev.} {\bf
  D96} (2017) 014001}, [\href{https://arxiv.org/abs/1704.03187}{{\tt
  1704.03187}}].

\bibitem{Aaron:2009aa}
{\scshape ZEUS, H1} collaboration, F.~D. Aaron et~al., \emph{{Combined
  Measurement and QCD Analysis of the Inclusive e+- p Scattering Cross Sections
  at HERA}}, \href{http://dx.doi.org/10.1007/JHEP01(2010)109}{\emph{JHEP} {\bf
  01} (2010) 109}, [\href{https://arxiv.org/abs/0911.0884}{{\tt 0911.0884}}].

\bibitem{Abramowicz:2015mha}
{\scshape ZEUS, H1} collaboration, H.~Abramowicz et~al., \emph{{Combination of
  measurements of inclusive deep inelastic ${e^{\pm }p}$ scattering cross
  sections and QCD analysis of HERA data}},
  \href{http://dx.doi.org/10.1140/epjc/s10052-015-3710-4}{\emph{Eur. Phys. J.}
  {\bf C75} (2015) 580}, [\href{https://arxiv.org/abs/1506.06042}{{\tt
  1506.06042}}].

\bibitem{Gribov:1972ri}
V.~N. Gribov and L.~N. Lipatov, \emph{{Deep inelastic e p scattering in
  perturbation theory}}, {\emph{Sov. J. Nucl. Phys.} {\bf 15} (1972) 438--450}.

\bibitem{Altarelli:1977zs}
G.~Altarelli and G.~Parisi, \emph{{Asymptotic Freedom in Parton Language}},
  \href{http://dx.doi.org/10.1016/0550-3213(77)90384-4}{\emph{Nucl. Phys.} {\bf
  B126} (1977) 298--318}.

\bibitem{Dokshitzer:1977sg}
Y.~L. Dokshitzer, \emph{{Calculation of the Structure Functions for Deep
  Inelastic Scattering and e+ e- Annihilation by Perturbation Theory in Quantum
  Chromodynamics.}}, {\emph{Sov. Phys. JETP} {\bf 46} (1977) 641--653}.

\bibitem{Lipatov:1976zz}
L.~N. Lipatov, \emph{{Reggeization of the Vector Meson and the Vacuum
  Singularity in Nonabelian Gauge Theories}}, {\emph{Sov. J. Nucl. Phys.} {\bf
  23} (1976) 338--345}.

\bibitem{Kuraev:1976ge}
E.~A. Kuraev, L.~N. Lipatov and V.~S. Fadin, \emph{{Multi - Reggeon Processes
  in the Yang-Mills Theory}}, {\emph{Sov. Phys. JETP} {\bf 44} (1976)
  443--450}.

\bibitem{Kuraev:1977fs}
E.~A. Kuraev, L.~N. Lipatov and V.~S. Fadin, \emph{{The Pomeranchuk Singularity
  in Nonabelian Gauge Theories}}, {\emph{Sov. Phys. JETP} {\bf 45} (1977)
  199--204}.

\bibitem{Balitsky:1978ic}
I.~I. Balitsky and L.~N. Lipatov, \emph{{The Pomeranchuk Singularity in Quantum
  Chromodynamics}}, {\emph{Sov. J. Nucl. Phys.} {\bf 28} (1978) 822--829}.

\bibitem{Gribov:1981ac}
L.~V. Gribov, E.~M. Levin and M.~G. Ryskin, \emph{{Singlet Structure Function
  at Small x: Unitarization of Gluon Ladders}},
  \href{http://dx.doi.org/10.1016/0550-3213(81)90007-9}{\emph{Nucl. Phys.} {\bf
  B188} (1981) 555--576}.

\bibitem{Gribov:1984tu}
L.~V. Gribov, E.~M. Levin and M.~G. Ryskin, \emph{{Semihard Processes in QCD}},
  \href{http://dx.doi.org/10.1016/0370-1573(83)90022-4}{\emph{Phys. Rept.} {\bf
  100} (1983) 1--150}.

\bibitem{Mueller:1985wy}
A.~H. Mueller and J.-w. Qiu, \emph{{Gluon Recombination and Shadowing at Small
  Values of x}},
  \href{http://dx.doi.org/10.1016/0550-3213(86)90164-1}{\emph{Nucl. Phys.} {\bf
  B268} (1986) 427--452}.

\bibitem{McLerran:1993ni}
L.~D. McLerran and R.~Venugopalan, \emph{{Computing quark and gluon
  distribution functions for very large nuclei}},
  \href{http://dx.doi.org/10.1103/PhysRevD.49.2233}{\emph{Phys. Rev.} {\bf D49}
  (1994) 2233--2241}, [\href{https://arxiv.org/abs/hep-ph/9309289}{{\tt
  hep-ph/9309289}}].

\bibitem{McLerran:1993ka}
L.~D. McLerran and R.~Venugopalan, \emph{{Gluon distribution functions for very
  large nuclei at small transverse momentum}},
  \href{http://dx.doi.org/10.1103/PhysRevD.49.3352}{\emph{Phys. Rev.} {\bf D49}
  (1994) 3352--3355}, [\href{https://arxiv.org/abs/hep-ph/9311205}{{\tt
  hep-ph/9311205}}].

\bibitem{JalilianMarian:1996xn}
J.~Jalilian-Marian, A.~Kovner, L.~D. McLerran and H.~Weigert, \emph{{The
  Intrinsic glue distribution at very small x}},
  \href{http://dx.doi.org/10.1103/PhysRevD.55.5414}{\emph{Phys. Rev.} {\bf D55}
  (1997) 5414--5428}, [\href{https://arxiv.org/abs/hep-ph/9606337}{{\tt
  hep-ph/9606337}}].

\bibitem{JalilianMarian:1997gr}
J.~Jalilian-Marian, A.~Kovner, A.~Leonidov and H.~Weigert, \emph{{The Wilson
  renormalization group for low x physics: Towards the high density regime}},
  \href{http://dx.doi.org/10.1103/PhysRevD.59.014014}{\emph{Phys. Rev.} {\bf
  D59} (1998) 014014}, [\href{https://arxiv.org/abs/hep-ph/9706377}{{\tt
  hep-ph/9706377}}].

\bibitem{Iancu:2000hn}
E.~Iancu, A.~Leonidov and L.~D. McLerran, \emph{{Nonlinear gluon evolution in
  the color glass condensate. 1.}},
  \href{http://dx.doi.org/10.1016/S0375-9474(01)00642-X}{\emph{Nucl. Phys.}
  {\bf A692} (2001) 583--645},
  [\href{https://arxiv.org/abs/hep-ph/0011241}{{\tt hep-ph/0011241}}].

\bibitem{Iancu:2001ad}
E.~Iancu, A.~Leonidov and L.~D. McLerran, \emph{{The Renormalization group
  equation for the color glass condensate}},
  \href{http://dx.doi.org/10.1016/S0370-2693(01)00524-X}{\emph{Phys. Lett.}
  {\bf B510} (2001) 133--144},
  [\href{https://arxiv.org/abs/hep-ph/0102009}{{\tt hep-ph/0102009}}].

\bibitem{Iancu:2003xm}
E.~Iancu and R.~Venugopalan, \emph{{The Color glass condensate and high-energy
  scattering in QCD}},  in \emph{In *Hwa, R.C. (ed.) et al.: Quark gluon
  plasma* 249-3363}.
\newblock 2003.
\newblock \href{https://arxiv.org/abs/hep-ph/0303204}{{\tt hep-ph/0303204}}.
\newblock \href{http://dx.doi.org/10.1142/9789812795533_0005}{DOI}.

\bibitem{Weigert:2005us}
H.~Weigert, \emph{{Evolution at small x(bj): The Color glass condensate}},
  \href{http://dx.doi.org/10.1016/j.ppnp.2005.01.029}{\emph{Prog. Part. Nucl.
  Phys.} {\bf 55} (2005) 461--565},
  [\href{https://arxiv.org/abs/hep-ph/0501087}{{\tt hep-ph/0501087}}].

\bibitem{Balitsky:1995ub}
I.~Balitsky, \emph{{Operator expansion for high-energy scattering}},
  \href{http://dx.doi.org/10.1016/0550-3213(95)00638-9}{\emph{Nucl. Phys.} {\bf
  B463} (1996) 99--160}, [\href{https://arxiv.org/abs/hep-ph/9509348}{{\tt
  hep-ph/9509348}}].

\bibitem{Kovchegov:1999yj}
Y.~V. Kovchegov, \emph{{Small x F(2) structure function of a nucleus including
  multiple pomeron exchanges}},
  \href{http://dx.doi.org/10.1103/PhysRevD.60.034008}{\emph{Phys. Rev.} {\bf
  D60} (1999) 034008}, [\href{https://arxiv.org/abs/hep-ph/9901281}{{\tt
  hep-ph/9901281}}].

\bibitem{Kovchegov:1999ua}
Y.~V. Kovchegov, \emph{{Unitarization of the BFKL pomeron on a nucleus}},
  \href{http://dx.doi.org/10.1103/PhysRevD.61.074018}{\emph{Phys. Rev.} {\bf
  D61} (2000) 074018}, [\href{https://arxiv.org/abs/hep-ph/9905214}{{\tt
  hep-ph/9905214}}].

\bibitem{GolecBiernat:2001if}
K.~J. Golec-Biernat, L.~Motyka and A.~M. Stasto, \emph{{Diffusion into infrared
  and unitarization of the BFKL pomeron}},
  \href{http://dx.doi.org/10.1103/PhysRevD.65.074037}{\emph{Phys. Rev.} {\bf
  D65} (2002) 074037}, [\href{https://arxiv.org/abs/hep-ph/0110325}{{\tt
  hep-ph/0110325}}].

\bibitem{GolecBiernat:1998js}
K.~J. Golec-Biernat and M.~Wusthoff, \emph{{Saturation effects in deep
  inelastic scattering at low $Q^2$ and its implications on diffraction}},
  \href{http://dx.doi.org/10.1103/PhysRevD.59.014017}{\emph{Phys. Rev.} {\bf
  D59} (1998) 014017}, [\href{https://arxiv.org/abs/hep-ph/9807513}{{\tt
  hep-ph/9807513}}].

\bibitem{Bartels:2002cj}
J.~Bartels, K.~J. Golec-Biernat and H.~Kowalski, \emph{{A modification of the
  saturation model: DGLAP evolution}},
  \href{http://dx.doi.org/10.1103/PhysRevD.66.014001}{\emph{Phys. Rev.} {\bf
  D66} (2002) 014001}, [\href{https://arxiv.org/abs/hep-ph/0203258}{{\tt
  hep-ph/0203258}}].

\bibitem{GolecBiernat:2006ba}
K.~J. Golec-Biernat and S.~Sapeta, \emph{{Heavy flavour production in DGLAP
  improved saturation model}},
  \href{http://dx.doi.org/10.1103/PhysRevD.74.054032}{\emph{Phys. Rev.} {\bf
  D74} (2006) 054032}, [\href{https://arxiv.org/abs/hep-ph/0607276}{{\tt
  hep-ph/0607276}}].

\bibitem{Iancu:2003ge}
E.~Iancu, K.~Itakura and S.~Munier, \emph{{Saturation and BFKL dynamics in the
  HERA data at small x}},
  \href{http://dx.doi.org/10.1016/j.physletb.2004.02.040}{\emph{Phys. Lett.}
  {\bf B590} (2004) 199--208},
  [\href{https://arxiv.org/abs/hep-ph/0310338}{{\tt hep-ph/0310338}}].

\bibitem{Kowalski:2003hm}
H.~Kowalski and D.~Teaney, \emph{{An Impact parameter dipole saturation
  model}}, \href{http://dx.doi.org/10.1103/PhysRevD.68.114005}{\emph{Phys.
  Rev.} {\bf D68} (2003) 114005},
  [\href{https://arxiv.org/abs/hep-ph/0304189}{{\tt hep-ph/0304189}}].

\bibitem{Rezaeian:2012ji}
A.~H. Rezaeian, M.~Siddikov, M.~Van~de Klundert and R.~Venugopalan,
  \emph{{Analysis of combined HERA data in the Impact-Parameter dependent
  Saturation model}},
  \href{http://dx.doi.org/10.1103/PhysRevD.87.034002}{\emph{Phys. Rev.} {\bf
  D87} (2013) 034002}, [\href{https://arxiv.org/abs/1212.2974}{{\tt
  1212.2974}}].

\bibitem{Rezaeian:2013tka}
A.~H. Rezaeian and I.~Schmidt, \emph{{Impact-parameter dependent Color Glass
  Condensate dipole model and new combined HERA data}},
  \href{http://dx.doi.org/10.1103/PhysRevD.88.074016}{\emph{Phys. Rev.} {\bf
  D88} (2013) 074016}, [\href{https://arxiv.org/abs/1307.0825}{{\tt
  1307.0825}}].

\bibitem{Albacete:2009fh}
J.~L. Albacete, N.~Armesto, J.~G. Milhano and C.~A. Salgado, \emph{{Non-linear
  QCD meets data: A Global analysis of lepton-proton scattering with running
  coupling BK evolution}},
  \href{http://dx.doi.org/10.1103/PhysRevD.80.034031}{\emph{Phys. Rev.} {\bf
  D80} (2009) 034031}, [\href{https://arxiv.org/abs/0902.1112}{{\tt
  0902.1112}}].

\bibitem{Albacete:2010sy}
J.~L. Albacete, N.~Armesto, J.~G. Milhano, P.~Quiroga-Arias and C.~A. Salgado,
  \emph{{AAMQS: A non-linear QCD analysis of new HERA data at small-x including
  heavy quarks}},
  \href{http://dx.doi.org/10.1140/epjc/s10052-011-1705-3}{\emph{Eur. Phys. J.}
  {\bf C71} (2011) 1705}, [\href{https://arxiv.org/abs/1012.4408}{{\tt
  1012.4408}}].

\bibitem{Lappi:2013zma}
T.~Lappi and H.~Mäntysaari, \emph{{Single inclusive particle production at
  high energy from HERA data to proton-nucleus collisions}},
  \href{http://dx.doi.org/10.1103/PhysRevD.88.114020}{\emph{Phys. Rev.} {\bf
  D88} (2013) 114020}, [\href{https://arxiv.org/abs/1309.6963}{{\tt
  1309.6963}}].

\bibitem{Iancu:2015joa}
E.~Iancu, J.~D. Madrigal, A.~H. Mueller, G.~Soyez and D.~N. Triantafyllopoulos,
  \emph{{Collinearly-improved BK evolution meets the HERA data}},
  \href{http://dx.doi.org/10.1016/j.physletb.2015.09.071}{\emph{Phys. Lett.}
  {\bf B750} (2015) 643--652}, [\href{https://arxiv.org/abs/1507.03651}{{\tt
  1507.03651}}].

\bibitem{Albacete:2012rx}
J.~L. Albacete, J.~G. Milhano, P.~Quiroga-Arias and J.~Rojo, \emph{{Linear vs
  Non-Linear QCD Evolution: From HERA Data to LHC Phenomenology}},
  \href{http://dx.doi.org/10.1140/epjc/s10052-012-2131-x}{\emph{Eur. Phys. J.}
  {\bf C72} (2012) 2131}, [\href{https://arxiv.org/abs/1203.1043}{{\tt
  1203.1043}}].

\bibitem{Ball:2017otu}
R.~D. Ball, V.~Bertone, M.~Bonvini, S.~Marzani, J.~Rojo and L.~Rottoli,
  \emph{{Parton distributions with small-x resummation: evidence for BFKL
  dynamics in HERA data}},  \href{https://arxiv.org/abs/1710.05935}{{\tt
  1710.05935}}.

\bibitem{Ewerz:2007md}
C.~Ewerz, A.~von Manteuffel and O.~Nachtmann, \emph{{On the Range of Validity
  of the Dipole Picture}},
  \href{http://dx.doi.org/10.1103/PhysRevD.77.074022}{\emph{Phys. Rev.} {\bf
  D77} (2008) 074022}, [\href{https://arxiv.org/abs/0708.3455}{{\tt
  0708.3455}}].

\bibitem{Ewerz:2011ph}
C.~Ewerz, A.~von Manteuffel and O.~Nachtmann, \emph{{On the Energy Dependence
  of the Dipole-Proton Cross Section in Deep Inelastic Scattering}},
  \href{http://dx.doi.org/10.1007/JHEP03(2011)062}{\emph{JHEP} {\bf 03} (2011)
  062}, [\href{https://arxiv.org/abs/1101.0288}{{\tt 1101.0288}}].

\bibitem{Bjorken:1970ah}
J.~D. Bjorken, J.~B. Kogut and D.~E. Soper, \emph{{Quantum Electrodynamics at
  Infinite Momentum: Scattering from an External Field}},
  \href{http://dx.doi.org/10.1103/PhysRevD.3.1382}{\emph{Phys. Rev.} {\bf D3}
  (1971) 1382}.

\bibitem{Stasto:2000er}
A.~M. Stasto, K.~J. Golec-Biernat and J.~Kwiecinski, \emph{{Geometric scaling
  for the total gamma* p cross-section in the low x region}},
  \href{http://dx.doi.org/10.1103/PhysRevLett.86.596}{\emph{Phys. Rev. Lett.}
  {\bf 86} (2001) 596--599}, [\href{https://arxiv.org/abs/hep-ph/0007192}{{\tt
  hep-ph/0007192}}].

\bibitem{James:1975dr}
F.~James and M.~Roos, \emph{{Minuit: A System for Function Minimization and
  Analysis of the Parameter Errors and Correlations}},
  \href{http://dx.doi.org/10.1016/0010-4655(75)90039-9}{\emph{Comput. Phys.
  Commun.} {\bf 10} (1975) 343--367}.

\bibitem{James:1994vla}
F.~James, \emph{{MINUIT Function Minimization and Error Analysis: Reference
  Manual Version 94.1}}, .

\bibitem{Frankfurt:1996ri}
L.~Frankfurt, A.~Radyushkin and M.~Strikman, \emph{{Interaction of small size
  wave packet with hadron target}},
  \href{http://dx.doi.org/10.1103/PhysRevD.55.98}{\emph{Phys. Rev.} {\bf D55}
  (1997) 98--104}, [\href{https://arxiv.org/abs/hep-ph/9610274}{{\tt
  hep-ph/9610274}}].

\bibitem{Beringer:1900zz}
{\scshape Particle Data Group} collaboration, J.~Beringer et~al., \emph{{Review
  of Particle Physics (RPP)}},
  \href{http://dx.doi.org/10.1103/PhysRevD.86.010001}{\emph{Phys. Rev.} {\bf
  D86} (2012) 010001}.

\bibitem{Aaron:2009af}
{\scshape H1} collaboration, F.~D. Aaron et~al., \emph{{Measurement of the
  Charm and Beauty Structure Functions using the H1 Vertex Detector at HERA}},
  \href{http://dx.doi.org/10.1140/epjc/s10052-009-1190-0}{\emph{Eur. Phys. J.}
  {\bf C65} (2010) 89--109}, [\href{https://arxiv.org/abs/0907.2643}{{\tt
  0907.2643}}].

\bibitem{Abramowicz:1900rp}
{\scshape ZEUS, H1} collaboration, H.~Abramowicz et~al., \emph{{Combination and
  QCD Analysis of Charm Production Cross Section Measurements in Deep-Inelastic
  ep Scattering at HERA}},
  \href{http://dx.doi.org/10.1140/epjc/s10052-013-2311-3}{\emph{Eur. Phys. J.}
  {\bf C73} (2013) 2311}, [\href{https://arxiv.org/abs/1211.1182}{{\tt
  1211.1182}}].

\bibitem{Andreev:2013vha}
{\scshape H1} collaboration, V.~Andreev et~al., \emph{{Measurement of inclusive
  $e p$ cross sections at high $Q^2$ at $\sqrt s =$ 225 and 252 GeV and of the
  longitudinal proton structure function $F_L$ at HERA}},
  \href{http://dx.doi.org/10.1140/epjc/s10052-014-2814-6}{\emph{Eur. Phys. J.}
  {\bf C74} (2014) 2814}, [\href{https://arxiv.org/abs/1312.4821}{{\tt
  1312.4821}}].

\end{thebibliography}

\providecommand{\href}[2]{#2}\begingroup\raggedright\endgroup

\end{document}